\documentclass[12pt]{article}

\usepackage[hmargin=2.4cm,vmargin=2.8cm]{geometry}
\usepackage{graphicx}
\usepackage{amsmath}
\usepackage{wasysym}
\usepackage{enumerate}
\usepackage{graphicx}  
\usepackage{dcolumn}   
\usepackage{bm}        
\usepackage{amssymb}   
\usepackage{enumerate}
\usepackage{amsmath}
\usepackage{float}
\usepackage{multibib}

\begin{document}

\title{Singularity stars}
\author { Ntina Savvidou\footnote{ksavvidou@upatras.gr} and Charis Anastopoulos\footnote{anastop@physics.upatras.gr}     \\
 {\small Department of Physics, University of Patras, 26500 Greece} }
\maketitle
\begin{abstract}
We study static spherically symmetric solutions  to  Einstein's equations with a repulsive singularity at the centre. We show that geodesics are extendible across the singularity, so the singularity  does not lead to pathological causality properties. It is best described as an irreducible spacetime boundary. As such it must be    assigned to an entropy, so that the total entropy
 is a sum of matter entropy and of singularity entropy. We evaluate the latter by using methods that have been developed for black hole thermodynamics, namely, Euclidean Quantum Gravity and Wald's Noether charge approach. Then, we use the maximum-entropy principle in order to show that regular solutions correspond to global maxima of the total entropy for stellar masses below the Oppenheimer-Volkoff limit, thus providing a thermodynamic justification to the regularity assumption employed in all stellar models. The maximum entropy principle also defines stable singular configurations for masses above the Oppenheimer-Volkoff limit, which we name singularity stars. We analyse their properties, and discuss the possibility that they correspond to a new type of astrophysical object that intermediates between neutron stars and black holes.
\end{abstract}


\section{ Introduction}
When the nuclear fuel of a star is exhausted, the star ejects a large part of its mass. The remnant settles to an equilibrium state and becomes a {\em compact star}. Compact stars correspond to stationary solutions of Einstein's equations. In particular, black holes are solutions to vacuum Einstein's equations; white dwarves and neutron stars are solutions to
 Einstein's equations with matter.

Ignoring rotation, a compact  star  is described by a static spherically symmetric metric
\begin{eqnarray}
ds^2  = - L^2(r) dt^2 + \frac{dr^2}{1-\frac{2m(r)}{r}} + r^2 (d\theta^2 +  \sin^2\theta d \phi^2) \hspace{0.2cm} \label{metric}
\end{eqnarray}
in terms of the lapse function $L(r)$, the mass function $m(r)$ and the standard coordinate system $(t, r, \theta, \phi)$.

Let the stellar surface be  located at $r = r_B > 0$. For $r \geq  r_B$, the metric Eq. (\ref{metric}) is of the Schwarzschild type, i.e.,  $m(r) = M $ and $L(r) = \sqrt{1-\frac{2M}{r}}$, where $M = m(r_B)$ is the stellar mass. For $r < r_B$, the metric satisfies the Tolman-Oppenheimer-Volkoff (TOV) equation
\begin{eqnarray}
\frac{dP}{dr}  = - \frac{(\rho +P) (m+4\pi r^3P)}{r^2(1-\frac{2m}{r})}, \label{TOV}
\end{eqnarray}
where $\rho$ is the density and $P$ is the pressure  of stellar matter. The TOV equation is supplemented by equations for the mass and lapse functions,
\begin{eqnarray}
\frac{d m}{dr} &=& 4 \pi r^2 \rho \label{dm}\\
\frac{1}{L} \frac{dL}{dr} &=& -\frac{1}{\rho +P} \frac{dP}{dr} \label{dN}.
\end{eqnarray}
In compact stars, no fuel is burning, hence, the relation between $P$ and $\rho$ can be expressed in terms of an {\em equilibrium} Equation of State (EoS). We assume that stellar matter consists of $k$   particle species. Then,  the density and the pressure are functions
$\rho = \rho(T, b_a)$ and $P = P(T, b_a)$
of  the local temperature $T$ and of the variables $b_a := \mu_a/ T$;  $\mu_a$ is the chemical potential associated to the $a$-th species ($a = 1, 2, \ldots, k$). This representation of the EoS is most convenient, because all $b_a$ are constants in the stellar interior at thermal equilibrium---see, Refs. \cite{KM75, SavAn14} and also the Appendix A. Then, Eq. (\ref{dN}) implies Tolman's law:
 $L T = T_{\infty}$,  where $T_{\infty}$ is a constant. $T_{\infty}$ coincides with the temperature of the stellar surface, as measured by an observer at infinity.

Solutions to Eqs. (\ref{TOV}---\ref{dN}), subject to the stellar boundary condition $P(r_B)=0$, are uniquely determined by the two variables $M$ and  $r_B$,  together with the constants $b_a$.  However,  the variables $(M, r_B, b_a)$ are not thermodynamically independent \cite{SavAn14}. As a result, not all solutions to Eqs.  (\ref{TOV}---\ref{dN}) are physical. Physical solutions are identified by an {\em additional} requirement. In the current theory of compact stars, this requirement is the {\em regularity condition}, namely, the assertion that the  spacetime geometry is  everywhere locally Minkowskian (i.e., regular).  This is possible only  if $m(0) = 0$.

Regular solutions have finite pressure $P_0$ at $r = 0$. Given $P_0$, Eqs. (\ref{TOV}---\ref{dN}) are integrated from the center outwards. The stellar surface $r = r_B$ is identified as the first zero of the pressure function $P(r)$;  the stellar mass is then defined as $M := m(r_B)$. It turns out that  there is an upper bound $M_{max}$ to the mass of  regular solutions, for any given EoS. The Chandrasekhar limit for white dwarves and the Oppenheimer-Volkoff  limit (OVL) for neutron stars are special cases of this bound. It is widely believed that  final states of gravitational collapse with   $ M > M_{max}$ are black holes---so far no proof exists.

The regularity condition leads to a successful theory for the structure of compact stars  but it is conceptually problematic.  Unlike the early days of general relativity, today we readily accept the physical relevance of solutions to Einstein's equations with singularities (i.e., black holes, the Big Bang), even if we agree that the singularities must somehow be removed/justified by a more fundamental theory. In what follows, we will show that the singular (i.e., non-regular) solutions of Eqs. (\ref{TOV}---\ref{dN})  have  no pathological behavior as far as  causality and predictability are concerned. Thus, their
  {\em a priori} exclusion  is {\em ad hoc}, and it requires further justification.

 Moreover, if General Relativity emerges as the macroscopic limit of a quantum gravity theory,  then geometry fluctuations (e.g., Wheeler's spacetime foam) should be taken into account. There is no reason why the fluctuations should not also include singular geometries, especially if such geometries are  well behaved. This is the case for the singular solutions of the TOV equations, because they do not involve any non-extendible causal geodesics. Hence, the stability of the regular solutions under  fluctuations  becomes an important issue.    In  thermodynamic systems, stability  is implemented  by the {\em maximum-entropy principle} (MEP), namely, the fact that the equilibrium configuration is the {\em global} maximum of entropy with respect to all variations of  unconstrained variables \cite{Call}.

The implementation of the MEP requires the assignment of gravitational entropy to the singularities. The singularity entropy is necessary for  thermodynamic consistency, but it also follows from the application of ideas and methods that originate from black hole thermodynamics. In the latter context,  Euclidean Quantum Gravity   \cite{EQG, GiHa77} and Wald's  Noether charge approach \cite{Wald93} were developed in order to define black hole entropy in general spacetimes. Both theories treat the black hole horizon as a  spacetime boundary and provide a rule  for evaluating  entropy to such boundaries.

We apply the methods above
   to   singular solutions of the TOV equations, treating the singularity at $r = 0$ as a spacetime boundary. We find  that the singularity must be assigned to an entropy, so that the   total entropy of a singular solution is a sum of matter entropy and singularity entropy. We derive the singularity entropy  up to a multiplicative constant $\lambda$. Then, we determine the value of this constant   by a universality argument and the
   comparison to a simple model where this idea was first implemented \cite{AnSav12}.

Having defined the entropy  functional, we prove that   the MEP leads to a first-principles justification of the regularity condition. Indeed, we  show that for $M \leq M_{max}$ the global maxima of entropy correspond to regular solutions. We also show that some regular solutions define local but not global maxima of entropy. These solutions define metastable states.

Furthermore, the MEP distinguishes physical solutions even for $M > M_{max}$: therefore {\em some} singular solutions  turn out to be thermodynamically stable. Since these solutions describe stars in which matter coexists with a singularity, we call them {\em singularity stars}.
Singularity stars follow from the very same set of equations that describe neutron stars, with no additional assumptions about the form of matter at high densities or the postulate of new gravity theories.  In their derivation, we employ only standard General Relativity, augmented by concepts from black hole thermodynamics.

The models we consider in this paper   ignore significant features of actual compact stars, such as rotation and magnetic fields. However, the results stated above are robust, in the sense that they follow from minimal assumptions, and do not depend on a specific choice of EoS. They appear as a generic feature of self-gravitating systems in thermodynamic equilibrium.
Keeping in mind the fact that the original prediction of compact stars came from equally simple models,  our results provide a good {\em prima facie} argument for the existence of singularity stars. A definite prediction requires further research. One needs to demonstrate that (at least some) thermodynamically stable singular solutions are also dynamically stable, and that singularity stars can arise as final states of gravitational collapse.

The structure of this paper is the following. In Sec. 2, we study the properties of singular solutions to the TOV and identify the singularity entropy. In Sec. 3, we implement the MEP for a simple equation of state in order to demonstrate the emergence of the regularity condition and the existence of singularity stars. In Sec. 4, we discuss the possibility that the singularity-star solutions correspond to physical stars.
  In Sec. 5, we summarize our results. The paper also contains three appendices. The appendix A summarizes general  properties of self-gravitating systems, the appendix B contains a detailed derivation of the singularity entropy from Euclidean Quantum Gravity, and in the appendix C we present further examples of our results in terms of various EoS.

\section{Thermodynamics  of singular solutions}
In  this section, we first examine the properties of singular solutions to the TOV equations. Then we construct the entropy function to used in the implementation of the MEP.

\subsection{Geometry of singular solutions}
We study solutions to the TOV equation (\ref{TOV}) and the supplementary equations (\ref{dm}) and (\ref{dN}). They define static, spherically symmetric metrics of the form Eq. (\ref{metric}), in terms of the lapse function $L(r)$ and the mass function $m(r)$.

First, we consider solutions with $m(0) = 0$. For finite $\rho_0 := \rho(0)$, Eq. (\ref{dm}) implies that in the vicinity of $r = 0$, $m(r) = \frac{4\pi\rho_0}{3} r^3$. Substituting into Eq. (\ref{dN}), we find that near $r = 0$, $L(r) = L_0 [1 + \frac{2 \pi r^2}{3} (\rho_0+3 P_0)]$, where $P_0 = P(0)$ and $L_0 = L(0)$.  The metric is locally Minkowskian everywhere, including $r = 0$. For this reason, these solutions are called regular.

Singular solutions correspond to non-zero values of $m(0)$. These solutions are less   known than regular ones. Some particular cases have been studied in Refs. \cite{ ZuPa84, AnSav12,  AnSav16, Kim17}. A positive value of $m(0)$ implies the existence of a horizon where $2m(r) = r$, for some value of $r$. However, it has been proven that the integration of the TOV equation from the boundary inwards does not encounter a horizon, provided some rather mild conditions are satisfied by the EoS \cite{ST}. Hence, no positive value of $m(0)$ is consistent with the existence of a stellar surface. For this reason, physical solutions that describe stars satisfy  $m(0) = - M_0$, for $M_0 > 0$. Substituting into Eq. (\ref{dN}), we find that near $r = 0$, $\frac{1}{L}\frac{dL}{dr} = - \frac{1}{2r}$; hence,
\begin{eqnarray}
L (r) = \frac{\eta}{\sqrt{r}},
\end{eqnarray}
for some constant $\eta > 0$. Around $r = 0$, the metric (\ref{metric}) becomes

\begin{eqnarray}
ds^2  = - \frac{\eta^2}{r} dt^2 + \frac{r dr^2}{2M_0} + r^2 (d\theta^2 + \sin^2\theta d \phi^2). \label{metric3}
\end{eqnarray}

By Tolman's law,  $T(r) \sim \sqrt{r}$, hence, $T(0) = 0$. In general, the limit $T\rightarrow 0$ may correspond to either zero or non-zero density and pressure, depending on whether the number density $n$ also goes to zero, or it remains finite. For example, in the free Fermi gas, the limit $T \rightarrow 0$ at constant $n$ leads to the well known expressions for degeneracy pressure and density. If the limits $T \rightarrow 0$ and $n\rightarrow 0$ are taken together, both density and pressure may vanish. Only the latter case   is compatible with the TOV equation near $r = 0$. To see this,   assume that either the density $\rho_0$ or the pressure $P_0$ at $r = 0$ is non-zero. Eq. (\ref{TOV}) implies that near $r =0$, $\frac{dP}{dr} = \frac{\rho_0 + P_0}{2r}$, hence, $P(r) = \frac{\rho_0 +P_0}{2} \log r +c$, for some constant $c$. As $P(0)$ diverges, we obtained a contradiction. Hence,   $\rho(0) = P(0) = 0$. The stress energy tensor vanishes at $r = 0$ and by Einstein's equations, so does the Ricci tensor $R_{\mu \nu}$.

The proper radius coordinate corresponding to the metric Eq. (\ref{metric3}) is $x = \frac{2}{3} r^{3/2}/\sqrt{2M_0}$. A two-sphere of proper radius $x$ around $r = 0$ has area equal to $4 \pi (\frac{9}{2} M_0)^{2/3} x^{4/3}$. This implies that the spacetime is not locally Minkowskian around $r = 0$, since in that case the area should be $4 \pi x^2$.  Thus, $r = 0$ manifests the characteristic behavior of   conical singularities.

Near $r = 0$, the metric Eq. (\ref{metric3}) behaves like a Schwarzschild solution with negative mass $-M_0$. In fact,   singular solutions at near-zero temperatures are locally approximated  by a negative-mass Schwarzschild solution within a ball of finite radius around the singularity---see, Sec. 3.2. In the Schwarzschild solution, scalars constructed from the Weyl tensor diverge as $r \rightarrow \infty$. Hence, the singularity at $r = 0$ is a curvature singularity.

 Nonetheless, the presence of this singularity does not lead to problems  causality and predictability in this spacetime. To see this, we consider
the geodesic equation near $r = 0$
\begin{eqnarray}
\dot{r}^2 = \frac{2M_0\epsilon^2}{\eta^2} - \frac{2M_0\sigma}{r} - \frac{2M_0 \ell^2}{r^3}, \label{geodes}
\end{eqnarray}
where $\epsilon$ and $\ell$ are constants. For causal geodesics they correspond to energy per unit mass and angular momentum per unit mass, respectively. The parameter $\sigma$ takes the value $1$ for timelike, $-1$ for spacelike  and $0$ for null geodesics. The dot denotes derivative with respect to an affine parameter $\lambda$.

No timelike geodesics arrive at the singularity. Infalling massive particles reach a minimal radius, $r_{min} = (\eta/\epsilon)^2$ and then they bounce back. Among null geodesics only  a set of measure zero reaches the singularity, namely, radial ones
 ($\ell = 0$). For these geodesics, $ \dot{r} = \pm \sqrt{2M_0}\epsilon/\eta $; hence, $r =  \pm \sqrt{2M_0}\epsilon/\eta \lambda + c$, for some constant $c$. The geodesics are extendible past the singularity: at $r = 0$,  radial incoming geodesics become  radial outgoing geodesics. Hence, all geodesics that start from past null infinity $\mathcal{I}^-$ reach the future null infinity $\mathcal{I}^+$.

Spacelike geodesics reach the singularity only if $\ell = 0$. In this case,  $ \frac{2M_0\epsilon^2}{\eta^2} $ is negligible in comparison to $\frac{2M_0\sigma}{r}$ near $r = 0$. Hence,
$r \dot{r}^2 = 2 M_0$, with solution $r = (\pm \frac{3}{2} \sqrt{2M_0} \lambda + c)^{2/3}$, for some constant $c$. Again, spacelike geodesics are extendible across the singularity.

It follows from the above that the singular solutions to the TOV equations involve no inextendible geodesics.   Causal geodesic completeness is a minimal condition for a spacetime to be considered singularity-free \cite{HawkingEllis}.  Observers outside the stellar surface will receive no non-causal signal originating from the singularity.

Next, we consider the case of accelerated observers reaching the singularity.  The key point is that all physical observers have finite proper acceleration; a rocket-ship moving towards the singularity can expend only a finite amount of fuel.

We evaluate the acceleration one-form for static observers near the singularity,
\begin{eqnarray}
a = \frac{d \log L}{dr} dr = -\frac{1}{2r} dr. \label{accel}
\end{eqnarray}
The minus sign in Eq. (\ref{accel}) implies a repulsive force.  The proper acceleration $\sqrt{a^{\mu}a_{\mu}}$ diverges like $r^{-1/2}$  as $r \rightarrow 0$.  This implies that in static configurations,  infinite pressure is required in order to push a material element towards the singularity.

Consider an infalling observer with four-velocity
 $u^{\mu} = (\dot{t}, \dot{r}, \dot{\theta}, \dot{\phi})$. Since $u^{\mu}u_{\mu} = -1$,
\begin{eqnarray}
\dot{t} = \frac{\sqrt{r}}{\eta} \sqrt{1 + \frac{r\dot{r}^2}{2M_0} + r^2 \dot{\theta}^2 + r^2 \sin^2\theta \dot{\phi}^2}. \label{dott0}
\end{eqnarray}
If $\dot{r}, \dot{\theta}$ and $\dot{\phi}$ are bounded, then
\begin{eqnarray}
\dot{t} \simeq \frac{\sqrt{r}}{\eta} \label{dottrr}
 \end{eqnarray}
 as $r \rightarrow 0$. Hence, $u^{\mu}$ approximates the four-velocity of a static observer. By Eq. (\ref{accel}), the observer requires infinite acceleration to reach the singularity.

Next, we examine the possibility that $\dot{r}, \dot{\theta}$ or $\dot{\phi}$ diverge as $r \rightarrow 0$. Since the tangential acceleration does not affect whether the observer reaches the singularity or not, we focus on  radially falling observers. If $\dot{r}$ diverges like $r^{-1/2}$ or more  slowly, Eq. (\ref{dottrr}) still applies and the earlier conclusion remains unchanged. If $\dot{r}$ diverges faster that $r^{-1/2}$, then by Eq. (\ref{dott0}),
\begin{eqnarray}
\dot{t} = \frac{r \dot{r}}{\eta \sqrt{2M_0}}.
\end{eqnarray}
Then the proper acceleration as $r \rightarrow 0$ is
\begin{eqnarray}
a^2:= a_{\mu}a^{\mu} = \frac{r\ddot{r}^2}{M_0}.
\end{eqnarray}
The proper acceleration is finite, only if  $|\ddot{r}|$ diverges at most with $r^{-1/2}$. Setting $\ddot{r} = c_1r^{-1/2}$ for some constant $c_1$, we obtain $\frac{1}{2}\dot{r}^2 +2c_1r^{1/2} = c_2$, where $c_2$ is a constant. Hence,  $\dot{r}$ is finite as $r \rightarrow 0$, contradicting our initial assumption.   We conclude that no observer with finite proper acceleration reaches the singularity. The spacetime is bounded-acceleration complete \cite{Ger68}.


To summarize, singular solutions to the TOV equations lead to no problems with causality and predictability. They satisfy geodesic completeness and bounded-acceleration completeness.
In Sec. 4.2,  the properties of the singularity at $ r = 0$ are further discussed in relation to the naked singularities that appear in gravitational collapse and to the cosmic censorship hypothesis

\subsection{Singularity entropy}

The  singularity at $r = 0$ defines a natural boundary of the spacetime manifold. We propose an entropy assignment to this boundary, in analogy   to the entropy of spacetime boundaries (horizons) in black hole thermodynamics. The rationale   is the following.

It is a long-standing belief among researchers in quantum gravity that spacetime singularities are removed or regularized by quantum gravity effects. In this scenario, general relativity applies as long as all curvature radii  remain significantly larger  than the Planck length. This reasoning applies to singular solutions to the TOV equations, where the Weyl curvature diverges as $r \rightarrow 0$. For these solutions, we introduce the length scale $\epsilon$, such that  the TOV equations apply  outside the sphere $\Sigma_{\epsilon}$, defined by $r = \epsilon$. Inside $\Sigma_{\epsilon}$, non-GR  degrees of freedom may become important, hence, leading to a spacetime geometry different from the
 one predicted by the TOV equation.

The key point is that an external, macroscopic   observer has no direct access to  the microscopic degrees of freedom inside $\Sigma_{\epsilon}$. Nonetheless, the observer can  describe those degrees of freedom thermodynamically,  by assigning  an entropy
  $S_{\epsilon}$ to the boundary $\Sigma_{\epsilon}$. This entropy is interpreted in the Boltzmann sense: it is the logarithm of microstates of all inaccessible microscopic degrees of freedom within $\Sigma_{\epsilon}$. If $\epsilon$ is much smaller than any macroscopic length scale that characterizes the geometry, $S_{\epsilon}$ is well approximated by its limiting value as $\epsilon \rightarrow 0 $, provided that the latter is finite. Hence, we are led to the definition  of singularity entropy
    $S_{sing}: = \lim_{\epsilon \rightarrow 0} S_{\epsilon}$.

A stronger argument for the existence of singularity entropy is provided by black hole thermodynamics.  Euclidean Quantum Gravity (EQG) \cite{EQG, GiHa77} and Wald's  Noether charge approach \cite{Wald93} are two methods that have been employed in order to define black hole entropy in general spacetimes. Both methods express black hole entropy as
 a
surface integral over the two-dimensional intersection of the horizon with a   Cauchy surface. In this sense,  the horizon is treated as a spacetime boundary. The key point is that both methods define entropy in a way that does not   apply specifically to   black hole spacetimes; in principle, they apply to all spacetime characterized by  Cauchy surfaces with boundary.  In particular, they apply to solutions of the TOV equation, and they lead to a derivation of the  singularity entropy $S_{sing}$.

EQG  is based on the properties of the (formal) gravitational path integral. For stationary solutions to Einstein's equations, it provides a relation between the Helmholtz free energy $F$  and  the Euclidean action $I_E$. Let $I[g, \phi]$ be the  action for  a metric $g$ and  matter fields $\phi$. For any  stationary solution $(g_0, \phi_0)$ to Einstein's equations,  we define the  Euclidean metric $g_0^E$  by substituting the Killing time $t$ with $-i \tau$, where $\tau$ is a periodic coordinate with period $\beta$.
 Then, the Helmholtz free energy for $(g_0, \phi_0) $ is
 $F = \beta^{-1} I_E[g^E_0, \phi_0]$,
where $I_E := -i I$.

In the Appendix B,  we evaluate  the Helmholtz free energy $F$ for   solutions to the TOV equation, both regular and singular. To this end, we consider an action functional  $I$ that is a sum of two terms $I_{grav}+I_{mat}$. The gravitational term $I_{grav}$ is the
 Gibbons-Hawking-York    action for spacetimes with boundary \cite{GiHa77, York}, and the matter term $I_{mat}$ is  the Kijowski-Smolski-Gornicka action  for gravitating perfect fluids \cite{ KSG91}. We find that
\begin{eqnarray}
F = M +T_{\infty} S_{mat} + \lambda \sqrt{2M_0} \eta. \label{helmh0}
 \end{eqnarray}
where  $S_{mat}$ is the entropy of matter and $\lambda$ is an undetermined dimensionless constant. EQG cannot determine the value of $\lambda$, because it depends on the explicit form of counter-terms in the gravitational action. In principle, $\lambda$ should be determined by a proper definition of the gravitational path integral.

 For  regular solutions, $M_0 = 0$. Then, Eq.
 (\ref{helmh}) reduces to the usual  thermodynamic identity for the Helmholtz free energy, $F = M +T_{\infty} S_{mat}$. For singular solutions, the total entropy $S_{tot} = (F -M)/T_{\infty}$ takes the form
 \begin{eqnarray}
 S_{tot} = S_{mat} + S_{sing},
 \end{eqnarray}
 where
 \begin{eqnarray}
S_{sing} =  \lambda \frac{\sqrt{2M_0} \eta}{T_{\infty}}  \label{noether}
 \end{eqnarray}
 is the entropy associated to the singularity.

  Thus, EQG   {\em verifies} that the singularity   contributes a term    $S_{sing}$    to the total entropy;  $S_{sing}$ is evaluated up to a multiplicative constant.

 Eq. (\ref{noether}) is confirmed by Wald's definition of gravitational entropy in terms of the Noether charge of spacetime diffeomorphisms \cite{Wald93}. Like  black-hole entropy, $S_{sing}$ is of the form $\frac{1}{T_{\infty}} Q(\xi)$, where
\begin{eqnarray}
Q(\xi) = \lim_{\epsilon \rightarrow 0} \frac{\lambda}{4 \pi} \oint_{\Sigma_{\epsilon}} d\sigma_{\mu \nu} \nabla^{\mu} \xi^{\nu} = \lambda \sqrt{2M_0}\eta
\end{eqnarray}
is the Noether charge associated to the timelike Killing vector $\xi = \frac{\partial}{\partial t}$. Like all charges,  $Q(\xi)$ is specified only up to a multiplicative constant $\lambda$.

A first-principles calculation of $\lambda$ requires a quantum theory of gravity. However, we can find its value through the following argument.
  If  $S_{sing}$ is of gravitational origin, it   depends only on the  spacetime geometry near the singularity, hence, only on the parameters $M_0$ and $\eta$.  In particular, $\lambda$ is the same for all singular solutions with the same $M_0$ and $\eta$, and it does not depend on the EoS. Therefore, it suffices to determine  $\lambda$ within a simple model system. This has been done in Ref. \cite{AnSav12}, where it was shown that a thermodynamically consistent description of self-gravitating radiation in a spherical box is possible  only if
\begin{eqnarray}
\lambda = -2.
\end{eqnarray}
 The negativity of $\lambda$  implies that the entropy of the singularity is smaller than the entropy of a locally Minkowskian spacetime. This is  expected,   otherwise  regular solutions would not be thermodynamically stable.

\section{Equilibrium states via the MEP}

 It is convenient to express the thermodynamic properties of gravitating systems in terms of the `free entropy' function, defined as
 \begin{eqnarray}
 \Omega := S + \sum_a b_a N_a.
  \end{eqnarray}
  $\Omega$ is  the Legendre  transform of the entropy $S$ with respect to the particle numbers $N_a$,  for the different particle species. For details, see Ref.  \cite{SavAn14} and also the Appendix A.

 For solutions to the TOV equation, $\Omega$ is a function of  the  mass $M$, the stellar radius $r_B$ and the constants $b_a$,
\begin{eqnarray}
\Omega(M, r_B, b_a) =   \int_0^{\infty}\frac{4 \pi  r^2 dr}{\sqrt{1-\frac{2m}{r}}} \frac{\rho +P}{T} -  \frac{2\sqrt{2M_0}\eta}{T_{\infty}}, \label{ometot}
\end{eqnarray}
where the entropy contribution from the singularity has been taken into account. According to the MEP, physical solutions maximize  the free entropy $\Omega$ with respect to $r_B$ for fixed $M$ and $b_a$.

\subsection{Equations of state}
 The implementation of the  MEP requires the choice of a specific  EoS. Our conclusions  are independent of this choice  as they   invariably follow from several different EoS. In the main text, we implement the MEP for  the  Oppenheimer and Volkoff EoS \cite{OV}  that describes a
single species of free fermions with mass $\mu$. Other  EoS  are treated  in the Appendix C.

Since we consider a single species of fermions, we drop the index $a$ in $b_a$ and $N_a$. The EoS for an ideal gas of free relativistic fermions is

\begin{eqnarray}
n &=& \frac{8D}{\mu} \left[ t^{3/2} F_{\frac{1}{2}}(t, b-t^{-1}) +  t^{5/2} F_{\frac{3}{2}}(t, b-t^{-1}) \right] \label{nff}\\
P &=& \frac{16D}{3} \left[ t^{5/2} F_{\frac{3}{2}}(t, b-t^{-1}) + \frac{1}{2} t^{7/2} F_{\frac{5}{2}}(t, b-t^{-1}) \right] \label{pff} \\
\rho &=& \mu n +  8D \left[ t^{5/2} F_{\frac{3}{2}}(t, b-t^{-1}) +  t^{7/2} F_{\frac{5}{2}}(t, b-t^{-1}) \right] \label{rff},
\end{eqnarray}
where $n$ is the number density, $P$ is the pressure,  and $\rho$ is the energy density; we also wrote  $t = T/\mu$ and $D = \frac{\mu^4}{8 \pi^2 \hbar^3}$. $F_{\alpha}$ stands for the  generalized Fermi integral
\begin{eqnarray}
F_{\alpha}(t, s) = \int_0^{\infty} dx \frac{x^{\alpha} \sqrt{2+ tx}}{e^{x - s}+1}. \label{finteg}
\end{eqnarray}

The regime of highly degenerate fermions corresponds to $t<< 1$ and $b>> 1$.  For $t<<1$ and $s <0$,  $F_{\alpha}(t, s)$ is  suppressed exponentially:  $F_{\alpha}(t, s) \sim e^{-|s|}$. Since, $s = b - t^{-1}$, in Eq. (\ref{nff}---\ref{rff}), it is convenient to employ the variable $Y :=  bt$. Then, $F_{\alpha} \sim e^{-b (Y^{-1}-1)}$ for $Y <1$.

For  $b >> 1$,
   $n, P$ and $\rho$ drop sharply as  $Y $  decreases from $Y > 1$ to  $Y< 1$; the width of the transition region is of order $b^{-1}$.  Hence, within an excellent approximation, we can set $n, P$ and $\rho$  to zero for $Y < 1$, while for $Y > 1$,
\begin{eqnarray}
F_{\alpha}(t,s) \simeq \int_0^{b-t^{-1}} x^{\alpha} \sqrt{2+ tx}.
\end{eqnarray}

The EoS  (\ref{nff}---\ref{rff}) for $Y > 1$ becomes
\begin{eqnarray}
n = \frac{D}{\mu} \nu(Y), \hspace{0.5cm}
P = D \upsilon(Y), \hspace{0.5cm}
\rho = D u(Y) ,
\end{eqnarray}
where
\begin{eqnarray}
\nu(Y) &=& \frac{8}{3} (Y^2 - 1)^{3/2} \\
\upsilon(Y) &=&  \frac{1}{3} Y\sqrt{Y^2-1} (2Y^2-5) +  \sinh^{-1}\sqrt{Y^2-1} \\
u(Y) &=&   Y\sqrt{Y^2-1} (2Y^2-1) -  \sinh^{-1}\sqrt{Y^2-1} .
\end{eqnarray}

The above expressions for $n, P$ and $\rho$ are standardly given as corresponding to $T = 0$ (and consequently to infinite $b$). Here, we view them as the dominant terms for small but non-vanishing temperature and large but finite $b$. Keeping $T$ non-zero and $b$ non-infinite is necessary for thermodynamic consistency.

The benefit of working in the near-zero temperature regime is that the free entropy  (\ref{ometot}) has non-trivial dependence on  only  two variables.

\subsection{Implementation of the MEP}
Next, we define the dimensionless variables $w = \sqrt{4\pi D}m$, and $x = \sqrt{4 \pi D} r$, and we express  Eqs. (\ref{TOV}---\ref{dN}) as
\begin{eqnarray}
\frac{dw}{dx} &=& x^2 u(Y)  \label{Yw1a}\\
\frac{dY}{dx} &=& - \frac{Y [w+x^3 \upsilon(Y)]}{x^2 (1 - \frac{2w}{x})}. \label{Yw2a}
\end{eqnarray}
We integrate Eqs. (\ref{Yw1a}---\ref{Yw2a}) from the boundary $x = x_B := \sqrt{4 \pi D} r_B$ inwards, with initial conditions   $Y(x_B) = 1$ (zero pressure at the stellar surface) and
  $w(x_B) = w_B$, where $w_B :=   \sqrt{4\pi D}M$. The local temperature at the stellar surface is  $\mu/b$, hence,
  \begin{eqnarray}
  T_{\infty} = \frac{\mu}{b} \sqrt{1 - \frac{2w_B}{x_B}}.
\end{eqnarray}

   The space $\Gamma$  of solutions to Eqs. (\ref{Yw1a}---\ref{Yw2a})   is two dimensional, each point specified by the dimensionless mass $w_B \in (0, \infty)$ and the dimensionless radius $x_B \in (0, \infty)$. The boundary of $\Gamma$ consists of the line $w_B = 0 $ that corresponds to Minkowski spacetime and the line $2 w_B = x_B$ that corresponds to  Schwarzschild black holes.

As explained in Sec. 2.1, points of $\Gamma$ correspond either to regular solutions with $w(0) = 0$, or to singular solutions with $w(0) < 0$. No point of $\Gamma$ corresponds to a spacetime with a horizon---i.e., to $w(0) > 0$---,because horizons are not encountered when integrating the TOV equations from the boundary inwards \cite{ST}.
Regular solutions define an one dimensional subspace of $\Gamma$ in the form of a spiral---see,  Fig. 1. The point $D$ of the spiral has  maximum mass $ w_{max}  \simeq 0.153$, and it defines the model's OVL.

\begin{figure}[tbp]
\includegraphics[height=5.5cm]{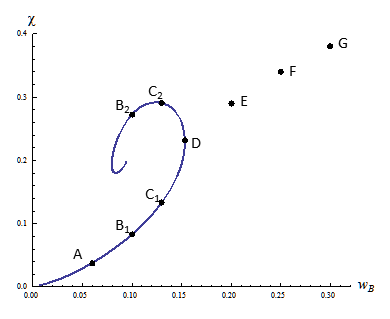} \caption{ \small  The spiral is the set of regular solutions that is defined by a functional relation between the ratio $\chi = 2w_B/x_B$ and the dimensionless mass $w_B$. Different regular solutions may have the same value of mass (e.g. points $B_1$ and $B_2$). The point $D$ corresponds to the maximum value of mass, $w_{max} \simeq 0.153$. Points $E, F$ and $G$ correspond to singular equilibrium solutions identified by the MEP---see, Fig. 4.a.}
\end{figure}

All points of $\Gamma$ outside the spiral correspond to singular solutions---see, Fig. 2 for a plot.
In a singular solution, $Y$ starts from the value $1$ at $x = x_B$ and increases with decreasing $x$. Near the singularity, $w$ becomes negative, so by Eq. (\ref{Yw2a}), $dY/dx > 0$, i.e., $Y$ decreases with decreasing $x$.  Eventually $Y(x_0) = 1$ at some point $x_0$. For $x < x_0$, $u$ and $\upsilon$ are proportional to $e^{-b(Y^{-1}-1)}$. They effectively vanish outside a region of width $b^{-1}$. The contribution of the intermediate region to the total energy and free entropy is negligible for $b >> 1$. We have verified this by solving the TOV equation with the full EoS (\ref{nff}---\ref{rff}).  Setting $u = v = 0$ for $x < x_0$ is an excellent approximation that becomes exact in the limit $b \rightarrow \infty$.
Hence,  the singularity is surrounded by a ball of dimensionless radius $x_0$ with no matter.

For $x < x_0$, $w$ is constant: $w = - w_0$, where $w_0 = \sqrt{4\pi D}M_0 $. The local geometry is that of a negative-mass Schwarzschild solution. Eq. (\ref{Yw2a}) implies that for $x < x_0$,
\begin{eqnarray}
Y = \sqrt{\frac{1+2w_0/x_0}{1+2w_0/x}}.
\end{eqnarray}

\begin{figure}[tbp]
\includegraphics[height=4.5cm]{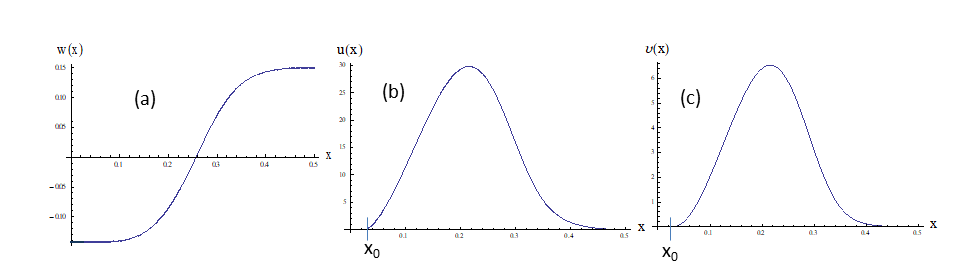} \caption{ \small  A singular solution to Eqs. (\ref{Yw1a}---\ref{Yw2a}) for $w_b = 0.15$ and $\chi = 0.15$. (a) The dimensionless mass $w$ is plotted as a function of the dimensionless radius $x$. (b) The dimensionless energy density $u$ is plotted as a function of $x$. (c) The dimensionless pressure $v$ is plotted as a function of $x$. Both pressure and density vanish for $x < x_0 \simeq 0.0266$. For $x < x_0$, the mass function is constant. Hence, the geometry is that of a negative-mass Schwarzschild solution. }
\end{figure}

Since $L = T_{\infty} b/(Y\mu)$,  we identify,
\begin{eqnarray}
\eta = \frac{T_{\infty} b}{\mu(4 \pi D)^{1/4}} \sqrt{\frac{2w_0}{1+\frac{2w_0}{x_0}}}
\end{eqnarray}

Then, the singularity entropy Eq. (\ref{noether}) becomes
\begin{eqnarray}
S_{sing} = - 4 \frac{b}{\mu \sqrt{4\pi D}} \frac{w_0}{\sqrt{1+\frac{2w_0}{x_0}}}.
\end{eqnarray}
$S_{sing}$  is straightforwardly evaluated for any solution after $x_0$ has been numerically determined.

The free entropy becomes
\begin{eqnarray}
\Omega(M, r_B, b) =   \frac{b}{\mu\sqrt{4\pi D}}H(w_B, x_B), \label{Hf}
\end{eqnarray}
where
\begin{eqnarray}
H(w_B, x_B) =  \int_{x_0}^{x_B} dx \frac{x^2 \nu(x)}{\sqrt{1 - \frac{2w(x)}{x}}} - 4  \frac{w_0}{\sqrt{1+\frac{2w_0}{x_0}}} \label{Omega2}
\end{eqnarray}
is a function on $\Gamma$.

 Since $\Omega$ is  proportional to $b$, we implement  MEP   by maximizing $H$ for constant $w_B$.  It is convenient to use the coordinates $(w_B, \chi)$ on $\Gamma$, where $\chi := 2w_B/x_B \in (0, 1)$ is the compactness parameter.

For $w_B < w_{max}$, the local maxima of the free entropy correspond to  regular solutions of the TOV equation. This is shown in Fig. 3.a, where
 $H(w_B, \chi)$  is plotted as a function of   $\chi$ for constant $w_B$.  For some  values of $w_B$, several local maxima exist.
 The maximum with the largest $\chi$ turns out to be the global maximum. Hence, it defines the equilibrium state. The other local maxima correspond to metastable states. Hence, the stable regular solutions correspond to the spiral segment from the origin to the point $D$ (Fig. 1), in agreement with the standard analysis of stellar stability.
  We conclude  that {\em  the regularity condition arises as a consequence of the MEP}.

The MEP also applies to solutions with $w_B > w_{max}$. The function  $H(w_B, \chi)$  has a unique maximum for each choice of $w_B$.  This maximum identifies a  {\em singular} solution as an equilibrium state---see, Fig. 3.b. Hence, there is a new phase of  thermodynamically stable  solutions to the TOV equation   with masses {\em higher than the OVL}. We  call these solutions  `singularity stars'.

We use the term `singularity stars' to refer only to equilibrium singular solutions, and not to all singular solutions of the TOV equations.  Equilibrium solutions define an one-dimensional submanifold of $\Gamma$---see,  Fig. 4.a. The segment of the curve from $w_B = 0$ to $w_B = w_{max}$ corresponds to regular equilibrium solutions and the segment of the curve from $w_{max}$ to infinity corresponds to singularity stars.

  \begin{figure}[]
\includegraphics[height=13.cm]{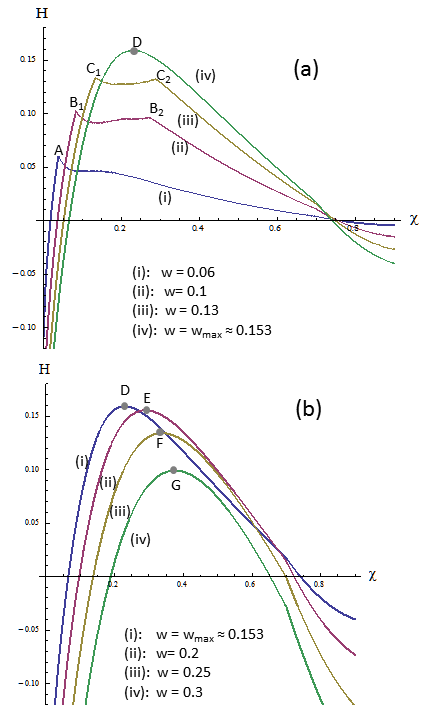} \caption{ \small  The normalized free entropy $H(w_B, x_B)$  of Eq. (\ref{Hf}) as a function of $\chi = 2w_B/x_B$ for constant $w_B$.  In Fig. 3.a,  $w_B \leq w_{max}$, and the global maxima are regular solutions.   In Fig. 3.b, $w_B \geq w_{max}$, and the global maxima are singular solutions. The entropy maxima are labeled by letters that correspond to the points of the $(w_B, \chi)$ plane in Fig. 1.
}
\end{figure}

 From a thermodynamic perspective, the OVL at $w_B = w_{max}$ is  the critical point for the transition between   the singularity-star phase and the regular-star phase. The first derivative of the function $\chi(w_B)$ for equilibrium solutions is discontinuous at the OVL. Quantities like the ratio $w_0/w_{max}$ and $|S_{sing}|$---plotted in Figs. 3.b and 3.c respectively---behave like order parameters: they vanish for the regular-star phase and are non-zero for the singularity-star phase. Hence,
the phase transition at the OVL is continuous\footnote{ Continuous phase transitions are characterized by universal critical exponents, which are straightforwardly evaluated in the present example. Unfortunately, they do not provide any information about the underlying microphysics.
Since the equilibrium configuration was determined by the MEP, all critical exponents are identical to the ones of Landau's theory, or equivalently of mean field theory. }.

  \begin{figure}[H]
\includegraphics[height=8cm]{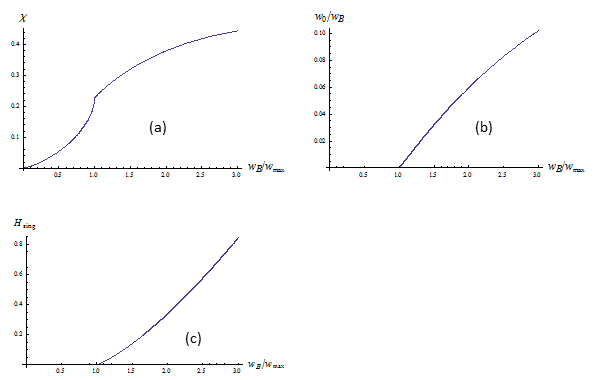} \caption{ \small  (a): the compactness parameter $\chi = 2w_B/x_B$ is plotted as a function of $w_B/w_{max}$ for equilibrium solutions, both regular and singular. The first part of the curve corresponds to the segment of the spiral in Fig. 1, from the origin to point $D$. (b): the ratio $w_0/w_B$ is plotted as a function of $w_B/w_{max}$. (c): the absolute value of the normalized singularity entropy $H_{sing} = \frac{\mu \sqrt{4\pi D}}{b}|S_{sing}|$ is plotted as a function of $w_B/w_{max}$.
}
\end{figure}

The above results  were faithfully reproduced by all physically meaningful EoS  that we tested. Some   examples are presented in   the Appendix C.
Only one restriction was found: the EoS should be such that
$
P \leq \frac{1}{3}\rho$
as $\rho \rightarrow \infty$, i.e., the pressure should not exceed the pressure of an ultra-relativistic ideal gas. One expects that this condition is satisfied by any asymptotically free quantum field theory. For an EoS with $P > \frac{1}{3}\rho$, regular solutions remain local maxima of the free entropy, but they are no longer global maxima. The free entropy blows up as
$\chi \rightarrow 1$. This suggests  that the only stable states compatible with such EoS are black holes.

\section{Can singularity stars exist?}
In this section, we discuss whether singular equilibrium solutions to the TOV equations correspond to actual astrophysical systems. A physically relevant solution must be (i) stable under perturbation and (ii) an end state  of a physical process with reasonable initial conditions.

\subsection{Stability}
 We will examine the stability of singularity stars under three types of fluctuations:  thermodynamical, dynamical and quantum.

 \subsubsection{Thermodynamic stability}
Singularity stars are identified by the MEP, so they satisfy  the most important criterion for thermodynamic stability. Being global entropy maxima, they cannot spontaneously evolve into a different static configuration.

In Fig. 5,  we plot the free entropy
 $H_{eq}$ of the equilibrium solutions as a function of the $w_B/w_{max}$.
 $H_{eq}$ is continuous across the
transition point $w_B/w_{max} = 1$, it increases for $w < w_1 \simeq 1.09 w_{max}$, and then it decreases to become negative at
 $w_- \simeq 2.5 w_{max}$. The same behavior characterizes the entropy of singularity stars that is derived from other EoS---see the Appendix C. In particular,   we found that $w_-/w_{max}$ always takes values between 2.5 and 3.

 Being negative, the free entropy of a singularity star with $w_B > w_-$ is always smaller than the free entropy of a black hole of the same mass. The latter equals the Bekenstein-Hawking entropy \cite{SavAn14}. Hence, singularity stars with $ w > w_-$ are unstable to decays towards  black holes.

  \begin{figure}[]
\includegraphics[height=5.5cm]{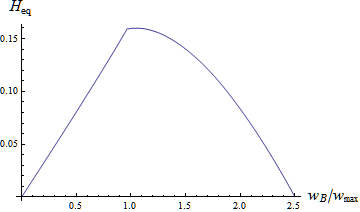} \caption{ \small  The normalized free entropy $H_{eq}$ of the equilibrium solutions is plotted as a function of the ratio $w_B/w_{max}$. Thermodynamic stability is manifested in the concavity of the curve. The free entropy is continuous across the transition point $w_B = w_{max}$, but its first derivative is discontinuous.
}
\end{figure}

The curve of the free entropy is everywhere concave, i.e., it satisfies
\begin{eqnarray}
\Omega(\lambda M_1 + (1-\lambda) M_2, b) \geq \lambda \Omega(M_1, b) + (1-\lambda) \Omega(M_2, b).
\end{eqnarray}
  Hence the set of equilibrium solutions (including singularity stars) satisfy another key criterion for thermodynamic stability.  Concavity guarantees that there is no energy flow between two identical systems in thermal contact, since $\Omega(M+\delta M, b) + \Omega (M-\Delta M, b) \leq 2 \Omega(M, b)$, i.e., the transfer of energy $\Delta M$ from one system to another is not favored.
In the regime considered here, $\Omega$ is proportional to $b$, and stability with respect to particle number holds trivially.

We conclude that singularity stars  $w_B \in(w_{max}, w_-)$ are   thermodynamically stable in the near-zero temperature regime. A generalization of the present results that takes into account finite-temperature and particle-mixing effects is necessary, in order to prove thermodynamic stability over the full thermodynamic state space.

\subsubsection{Dynamical stability}
 A stationary solution to  Einstein's equations is (linearly) dynamically stable, if no linearized perturbation admits runaway solutions. Here, we discuss radial perturbations that are typically characterized by the strongest instabilities.

 We consider spherically symmetric metrics of the form
\begin{eqnarray}
ds^2 = - L^2(r,t)dt^2 + \frac{dr^2}{1-\frac{2m(r, t)}{r}} + r^2 (d\theta^2 + \sin^2\theta d \phi^2),
\end{eqnarray}
with time-dependent lapse $L(r, t)$ and mass function $m(r, t)$ that perturb a static (equilibrium) solution to Einstein's equations. The linearized perturbations to Einstein's equation can be expressed in terms of an appropriately chosen function $f(r, t)$ that satisfies a hyperbolic equation
\begin{eqnarray}
W\ddot{f} = (Sf')' +Qf \label{pert1}
\end{eqnarray}
where $S, Q$ and $W$ are functions of $r$ that are determined by the equilibrium solution. A dot denotes differentiation with respect to $t$ and a prime denotes differentiation with respect to $r$.

For oscillatory perturbations  $f(r) = \zeta_{\omega}(r)e^{-i \omega t}$ with frequency $\omega$, Eq. (\ref{pert1}) becomes
\begin{eqnarray}
 (S\zeta_{\omega}')' +Q\zeta_{\omega} +\omega^2 W \zeta_{\omega}= 0.\label{pert2}
\end{eqnarray}
Eq. (\ref{pert2}) is a Sturm-Liouville equation with eigenvalues $\omega^2$. Assuming appropriate boundary conditions at $r = 0$ and $r = r_B$,  all eigenvalues $\omega_0^2, \omega_1^2, \dots$ are real, and they are ordered as $\omega_0^2 < \omega_1^2 < \ldots < \omega_n^2 < \ldots \rightarrow \infty$.   A negative eigenvalue $\omega^2$ signifies a mode growing unboundedly; hence, dynamical instability.  It follows that   $\omega_0^2 > 0$ is  a necessary and sufficient condition for dynamical stability. This condition can also be expressed as
\begin{eqnarray}
\int_0^{r_B}dr (S\zeta'^2 - Q\zeta^2) > 0, \label{condsta}
\end{eqnarray}
for {\em any} twice differentiable function $\zeta$ subject to the appropriate boundary conditions \cite{MTW}.

For regular solutions to the TOV equation, one typically assumes that perturbations preserve regularity, i.e., that the mass function at the origin always vanishes so that $\delta m = 0$ \cite{Yabu73}. This boundary condition is obviously inappropriate for singular solutions. Since all singular solutions satisfy  $P(0) = \rho(0) = 0$, the obvious requirement is that perturbations of density and pressure vanish at $r = 0$: $\delta \rho(0, t) = \delta P(0, t) = 0$. Physical intuition obtained from the stability analysis of regular solutions is not applicable to singular solutions. For example, the common conclusion that large values of the central pressure $P(0)$ lead to instability does not apply to singular solutions.

 Stability is a global property of the spacetime geometry.
Different metrics   lead to different functions $S, Q$ and $W$ in Eq. (\ref{pert2}); hence, to  different sets of eigenvalues. This applies in particular to metrics that coincide in  a finite region but strongly differ everywhere else, like the negative-mass Schwarzschild spacetimes and singular solutions to the TOV equation. Hence,
 the known instability of negative-mass Schwarzschild spacetimes \cite{negsch} is irrelevant to the stability problem of singularity stars.

As a matter of fact, the singularity at $r = 0$   {\em enhances} the stability   of equilibrium solutions. For adiabatic, radial perturbations,  the function $Q(r)$ of Eq. (\ref{pert2}) is \cite{MTW}
\begin{eqnarray}
Q(r) = \frac{L^{3/2}(r)}{\sqrt{1 - \frac{2m(r)}{r}}}\left[ \frac{[P'(r)]^2}{r^2[\rho(r) + P(r)]} - \frac{4 P'(r)}{r^3} + 8 \pi  \frac{[\rho(r) + P(r)] P(r)}{r^2 \left(1 - \frac{2m(r)}{r}\right)} \right].
\end{eqnarray}
where $L(r), m(r), P(r)$ and $\rho(r)$ refer to the equilibrium solution.

By Eq. (\ref{condsta}),  positive values of $Q$  contribute towards instability. For regular solutions, $Q(r) > 0$ for all $r$. However, for singular solutions,  $Q(r)$ becomes negative near $r = 0$,
\begin{eqnarray}
Q(r) \simeq - \frac{7 \eta^{3/2}[\rho(r) + P(r)]}{4\sqrt{2M_0} r^5} < 0.
\end{eqnarray}
Since $S(r) > 0$, the region around the singularity always contributes a positive term to the integral (\ref{condsta}). The singularity tends to stabilize the equilibrium solutions. This makes good physical sense: the repulsion prevents perturbations that cause   the inner layers of the star to collapse.



\subsubsection{Quantum stability}
Quantum instabilities occur when the  ADM- Hamiltonian operator for asymptotically flat spacetime geometries admits  negative eigenvalues. In this case, the Minkowski spacetime is quantum mechanically unstable because it can decay to a lower energy state. Lacking a quantum theory of gravity, the Hamiltonian cannot be constructed directly.  Studies of quantum instability typically look for runaway solutions in  hybrid quantum-classical dynamics,  like semi-classical gravity or the use of the quantum effective action for deriving classical equations of motion.

Quantum instabilities may be important for vacuum spacetimes, where quantum fluctuation are the primary contribution to energy---see, for example the discussion of the quantum stability of Minkowski spacetime \cite{Hor80, AMM03, HRV04}. However, they are negligible for the non-vacuum solutions studied in this paper, since the latter are characterized by ADM mass with macroscopically large (in fact, astronomical) values. Hence,  quantum instabilities are not relevant to singularity stars.

Quantum fluctuations may only become important at the limit where the compactness parameter $\chi$ tends to unity. Ref. \cite{AnSav16} suggests that quantum effects induce additional terms to the entropy functional that may affect the transition from the
  neutron-star  and/or singularity-star phase to the black hole phase.

\subsection{Formation of singularity stars}

Singularity stars are acceptable physical systems, only if they constitute an end-state of the full Einstein equations for generic initial conditions.
In this context,  ``generic"  means that the set of initial data leading to the formation of a singularity star must  be of non-zero measure in the set of all initial data.

Thermodynamics suggests several possible ways for the formation of singularity stars. For example, one may consider the continuous absorbtion of radiation by a neutron star  until its mass crosses the OVL; or the merging of two neutron stars towards an end-state with mass above the OVL. Both processes are entropically allowed and they  require no particular fine-tuning in their initial conditions. They should lead to the formation of singularity stars, provided that the latter are stable.

Of course, the most important mechanism for the formation of compact stars is gravitational collapse. The formation of spacetimes with naked singularities is a common feature in gravitational collapse models. In particular, spherically symmetric gravitational collapse leads to the formation  of naked singularities for generic  (spherically symmetric)  initial conditions \cite{ES79, Ch84}. The detailed properties of the naked singularities are model-dependent, and there is a long-standing discussion about their physical relevance and stability---see,  \cite{Joshi1, Joshi2} and references therein.

Given the above, the emergence of singularity stars as asymptotic states of gravitational collapse is a plausible assumption. Indeed,  solutions of the TOV equations with naked singularities surrounded by matter have already been obtained as end-states of gravitational collapse. However,  existing studies have focussed on toy models with simplified matter content (purely tangential pressure) \cite{JMN11, JMN13}, so they have not produced the characteristic repulsive singularity found here.

A thermodynamically consistent EoS for matter is essential for the derivation of singularity stars as asymptotic states of gravitational collapse. Barytropic EoS  that are often employed in such models are not compatible with the low density regime around the singularity. In particular, linear equations of state $P = w \rho$, for  $w \in (0, 1)$, are not compatible with the formation of a stellar surface. Polytropic equations of state $P = k \rho^{\gamma}$, for  $k> 0$ and $\gamma > 1$, lead by Eq. (\ref{dpt}) to  a relation between temperature and density of the form
\begin{eqnarray}
T = C(1 +k\rho^{\gamma -1})^{\frac{\gamma}{\gamma -1}},
\end{eqnarray}
where $C> 0$ is an integration constant. Hence, $T$ cannot be made smaller than $C$ at any density. This  behavior  is incompatible with the singular solutions to the TOV equation of Sec. 2.1 where $T \rightarrow 0$ near the singularity. The derivation of singularity stars as asymptotic states of gravitational collapse requires an EoS that remains consistent at all ranges of temperature and density. To the best of our knowledge,  this has not yet been done.

We also note that singularity stars, defined in terms of entropy maximization,  can only be arrived at by entropy-increasing processes. Even with the appropriate EoS, adiabatic collapse models cannot distinguish between singularity stars and non-entropy-maximizing singular solutions to the TOV equation.
Therefore, it is necessary to incorporate  thermodynamically irreversible processes, such as  dissipation and heat transfer in the description of collapsing matter---see, for example, Ref. \cite{HeSa04}.

The naked singularities considered here do not involve non-extendible geodesics, and in this sense they are much milder than the ones appearing in existing models of gravitational collapse. In fact, they may turn out to be compatible with
  the {\em weak cosmic censorship hypothesis} (WCCH) \cite{Pen69}. The usual formulation of the WCCH asserts that the  maximal Cauchy
evolution of asymptotically flat, regular  data is an asymptotically flat spacetime with complete future null infinity $\mathcal{I}^+$. In Sec. 2.1, we showed that in singular solutions to the TOV equation, no null geodesics terminate at the singularity and only a set of measure zero among null geodesics passes $r = 0$. Hence,  $\mathcal{I}^+$ is complete in the equilibrium solution. Whether it remains complete in the full collapsing spacetime is an open issue. However, we expect that
 any violations of the WCCH during the collapse process will be  much less severe than in existing models of gravitational collapse.

\subsection{Summary}
We  argued that  the existence of singularity stars cannot be ruled out on the basis of existing knowledge.  The results of this paper make a good prima facie case for singularity stars, but
a definitive prediction requires  further research. In particular, it is necessary to (i)
   construct the full thermodynamic space of singularity stars, in order to identify all regimes of thermodynamic stability; (ii) prove  dynamical stability for generic singularity-star solutions; and (iii)
 show that singularity stars emerge  as asymptotic states of gravitational collapse.

\section{Conclusions}
The main results of this paper are the following.

 First, we argued that the singularities in solutions of the TOV equation ought to be assigned entropy, and we  evaluated this entropy using EQG. This result provides an explicit confirmation of Penrose's conjecture about the relation between singularities and entropy \cite{Pen1}. It also demonstrates the usefulness of the EQG in identifying gravitational contributions to entropy in context other than black hole physics. The  issue arises whether the  singularity entropy derived here can be reproduced by existing quantum gravity theories.

 Second,  we used the entropy assigned to the singularity in order to show  that  regular solutions to the TOV equation satisfy the MEP for general EoS. Hence, we provide a physical justification for the regularity condition that has so far been assumed {\em a priori}. We also showed how the MEP distinguishes between thermodynamically stable and metastable regular solutions.

  Third, we showed that the    MEP also identifies singular equilibrium solutions with masses higher than the OVL. Hence, the thermodynamics of self-gravitating systems involves a singularity-star phase in addition to the regular-star phase.
     The prediction of singularity stars is robust, as it involves no conjectures about the behavior of nuclear matter at higher densities, exotic forms of matter, or dynamical modifications of General Relativity.

Finally, we  argued that the existence of singularity stars cannot be ruled out on the basis of known physics, and that further work is required in order to establish whether they correspond to genuine astrophysical objects or not.


\vspace{1cm}

\appendix

\section{Thermodynamic description of self-gravitating systems}
In this section, we summarize some thermodynamic properties of self-gravitating systems in equilibrium \cite{SavAn14}.

 We  consider a static globally hyperbolic spacetime $M = {\pmb R}\times \Sigma$ with four-metric
\begin{eqnarray}
ds^2 = -L^2(x) dt^2 + h_{ij}(x) dx^i dx^j, \label{4metric}
\end{eqnarray}
expressed in terms of the spatial coordinates $x^i$ and the time coordinate $t$.  $L$ is the lapse function, and $h_{ij}$ is a $t$-independent Riemannian three-metric on the surfaces $\Sigma_t$ of constant $t$. The time-like   unit normal on $\Sigma_t$ is $n_{\mu}= L \partial_{\mu}t$ and the extrinsic curvature tensor on $\Sigma_t$ vanishes.

 Let  $C \subset \Sigma$ be a compact spatial region, with boundary $B = \partial C$. $C$ contains an isotropic fluid in thermal and dynamical equilibrium, described by the stress-energy tensor
\begin{eqnarray}
T_{\mu \nu} = \rho n_{\mu}n_{\nu} + P (g_{\mu \nu} + n_{\mu} n_{\nu}), \label{tumn}
\end{eqnarray}
where $\rho(x)$ and $P(x)$ are the energy density and the pressure, respectively.

 The continuity equation $\nabla_{\mu}T^{\mu \nu} = 0$ for the metric (\ref{4metric}) becomes
\begin{eqnarray}
\frac{\nabla_i P}{\rho +P} = - \frac{\nabla_iL}{L}. \label{cont}
\end{eqnarray}

We assume that the fluid consists of $k$  particle species. The associated particle-number densities $n_a(x)$, $a = 1, \ldots, k$, together with the energy density $\rho(x)$ define the thermodynamic state space. All local thermodynamic properties of the fluid are encoded in the entropy-density functional $s(\rho, n_a)$. The first law of thermodynamics takes the form
\begin{eqnarray}
T ds = d \rho - \sum_a \mu_a dn_a, \label{1st}
\end{eqnarray}
where $\mu_a = - T \frac{\partial s}{\partial n_a}$ is the chemical potential associated to particle species $a$ and $T = \left(\frac{\partial s}{\partial \rho}\right)^{-1}$ is the local temperature. The pressure $P$ is defined through the Euler equation
\begin{eqnarray}
\rho + P - Ts -\sum_a \mu_a n_a = 0. \label{euler}
\end{eqnarray}
 Combining  Eqs. (\ref{euler}) and  (\ref{1st}), we derive the Gibbs-Duhem relation,
$dP  = s dT + \sum_a n_a d\mu_a$.

We maximize the total entropy of matter $S = \int_C d^3x \sqrt{h} s(\rho, n_a)$ for fixed values of the total particle numbers in $C$, $N_a = \int_C d^3x \sqrt{h} n_a $. To this end, we vary the function
  \begin{eqnarray}
\Omega =  S + \sum_a b_a N_a
  \end{eqnarray}
 with respect to $n_a$, where  $b_a$ are Lagrange multipliers. $\Omega$ is a Massieu function obtained by the Legendre transform of entropy, we will refer to it as the {\em free entropy} of the system.

 Variation with respect to $n_a$
\begin{eqnarray}
 \delta \Omega =   \sum_a \int_C d^3 x \sqrt{h} \left( -\frac{\mu_a}{T} + b_a\right) \delta n_a = 0,
\end{eqnarray}
leads to
 $b_a =  \frac{\mu_a}{T}$.  Hence, for equilibrium configurations the thermodynamic variables $\frac{\mu_a}{T}$ are constant in $C$.

We introduce the free entropy density $\omega$  as the Legendre transform of the entropy density $s$ with respect to $n_a$
\begin{eqnarray}
\omega(\rho, b_a) := s -\sum_a \frac{\partial s}{\partial n_a} n_a= s + \sum_a b_a n_a = \frac{\rho + P}{T}. \label{omega}
\end{eqnarray}
so that
\begin{eqnarray}
\Omega = \int_C d^3x \sqrt{h} \omega(\rho,b_a). \label{OM}
\end{eqnarray}

Substituting Eq. (\ref{omega}) into Eq. (\ref{1st}), we obtain
\begin{eqnarray}
 d \omega = \frac{d \rho}{T} + \sum_a n_a db_a.
\end{eqnarray}
It follows  that
$ T^{-1} = \partial \omega/\partial \rho$ and  $n_a =  \partial \omega/\partial b_a$. The Gibbs-Duhem relation becomes
$dP = \omega dT + T \sum_a n_a db_a$.

For entropy-maximizing configurations, $db_a = 0$, hence,
 \begin{eqnarray}
\frac{ dP}{dT} = \omega = \frac{P+\rho}{T}. \label{dpt}
 \end{eqnarray}
  Combining with Eq. (\ref{cont}), we obtain
\begin{eqnarray}
\frac{\nabla_iT}{T} = - \frac{\nabla_i L}{L}, \label{tol1}
\end{eqnarray}
which leads to Tolman's relation between local temperature and lapse function
\begin{eqnarray}
L T = T_{\infty}, \label{tolman}
\end{eqnarray}
where $T_{\infty}$ is    the temperature  seen by an observer at infinity (where $L = 1$).

\section{Singularity entropy from Euclidean quantum gravity}
In Euclidean Quantum Gravity, entropy and other thermodynamic variables are defined in terms of the Euclidean action associated to solutions of the classical equations of motion. In what follows, we evaluate the action functional for  general solutions to the TOV equations.
 \subsection{The action for a self-gravitating perfect fluid}
 Relativistic hydrodynamics can be formulated as a Lagrangian field theory, so that the corresponding Euler-Lagrange equations coincide with Einstein's equations for a perfect fluid. Here, we follow the formulation of Kijowski, Smolski and Gornicka \cite{KSG91}.

 We consider a globally hyperbolic spacetime, described by a Lorentzian metric $g$ and a spacetime manifold $M$ with topology ${\pmb R} \times \Sigma$. The field variables for the fluid are maps $\zeta: M \rightarrow {\pmb R} \times Z$, represented by four scalar fields $\zeta^0(x), \zeta^1(x), \zeta_2(x), \zeta^3(x)$. $Z$ is the {\em matter space}, its points correspond to material particles.

 We define the particle current  $j^{\mu} = \epsilon^{\mu\nu\rho \sigma} \zeta^1_{\nu} \zeta^2_{\rho} \zeta^3_{\sigma}$, where $\zeta^i_{\nu} = \partial \zeta^i/\partial x^{\nu}$ for $i = 1, 2, 3$. We write  $j^{\mu} = n u^{\mu}$, where $n = \sqrt{-j_{\mu}j^{\mu}}$ is the particle number density and $u^{\mu}$ is the unit four-vector of fluid velocities. We also express the local temperature $T$ as a function of $\zeta^0$, as $T = u^{\mu} \partial_{\mu} \zeta^0$.

Next, we consider  the action functional
\begin{eqnarray}
I_{mat}[g, \zeta] = - \int d^4x \sqrt{-g} f(n, T), \label{Imat}
\end{eqnarray}
 where $f(n, T)$ is the Helmholtz free energy density. Variation with respect to $\zeta^0$ and $\zeta^i$ leads to the conservation equation
 \begin{eqnarray}
 \nabla_{\mu} T^{\mu \nu} = 0,
 \end{eqnarray}
 for the perfect-fluid stress-energy tensor
 \begin{eqnarray}
 T_{\mu \nu} =  \rho u_{\mu} u_{\nu} + (g_{\mu \nu} + u_{\mu} u_{\nu})P \label{tmn}
 \end{eqnarray}
 and to the conservation equation for entropy
 \begin{eqnarray}
 u^{\mu} \partial_{\mu}s =0.
 \end{eqnarray}
 The pressure $P$ and the entropy density $s$ are defined in terms of derivatives of $f$: $P = n^2 (\partial f/\partial n)_T$ and $s = - (\partial f/\partial T)_n$.

 A self-gravitating fluid on a spacetime $M$ with a boundary $\partial M$ is described by the action
  \begin{eqnarray}
  I[g, \zeta] = \frac{1}{16} \pi \int d^4x \sqrt{-g}R -  \int d^4x \sqrt{-g} f(n, T) + \frac{1}{8 \pi} \int_{\partial M} d^3y\sqrt{|\gamma|} (K-K_{ref}). \label{action}
  \end{eqnarray}
 The boundary term in the action (\ref{action}) is included so that the variational principle is well-defined \cite{York, GiHa77};
   $y$ are coordinates in $\partial M$, $\gamma$ is the induced metric on $\partial M$ and $K$ is the trace of the extrinsic curvature at the boundary.  $K_{ref}$ is the trace of the extrinsic curvature of $\partial M$ with respect to a reference metric. The term containing $K_{ref}$  is included in Eq. (\ref{action}) in order to make the total action finite.

  Variation with respect to the metric leads to Einstein's equations $G_{\mu \nu} = 8 \pi T_{\mu \nu}$, where $T_{\mu \nu}$ is given by Eq. (\ref{tmn}). For solutions to the equations of motion $R = - 8 \pi T_{\mu \nu}g^{\mu \nu} = 8\pi (-\rho +3 P)$, and the action (\ref{action}) becomes
  \begin{eqnarray}
I[g, \zeta] = -\frac{1}{2}\int_M d^4x \sqrt{-g}  (\rho +3 P)  - \int_M d^4x \sqrt{-g} Ts      + \frac{1}{8 \pi} \int_{\partial M} \sqrt{|\gamma|} (K - K_{ref}). \label{acsg2}
\end{eqnarray}
 For a static metric of the form (\ref{4metric}),
  Einstein's equations imply that $4 \pi (\rho+3P) L = \nabla_i\nabla^iL$,
  where $\nabla_i$ is the covariant derivative on the three-surface  of constant $t$. Hence, the first term of Eq. (\ref{acsg2}) becomes  $-\frac{1}{8 \pi}\int_M d^4x \sqrt{h}   \nabla_i\nabla^iL$, where $h$ is the determinant of the three-metric $h_{ij}$. Using Tolman's law, the second term in Eq. (\ref{acsg2}) becomes $- T_{\infty} S_{mat} \int dt$, where $S_{mat}$ is the total entropy of the fluid at a single moment of time
  \begin{eqnarray}
  S_{mat} =  \int_{\Sigma}d^3x  \sqrt{h} s
  \end{eqnarray}

 \subsection{The action for singular solutions}
Next, we evaluate the action function to static spherically symmetric solutions to Einstein's equations. The associated metric is given by
Eq.  (\ref{metric}). The star's surface is defined by the condition $r = r_B$. For $r > r_B$, the solution is Schwarzschild, $ m(r) = m(r_B) = M$ and
$L(r) = \sqrt{1-\frac{2M}{r}}$. For $r < r_B$, $L$ and $m$ are  solutions to the TOV equation.

The spacetime has topology ${\pmb R} \times \Sigma$. The spatial manifold $\Sigma$ is bounded by two spheres
$C_0$ and $C_1$, defined by the conditions $r = r_0$ and $r = r_1$ respectively; $C_0$ is inside the star and $C_1$ is outside.
Eventually we will take the limits $ r_0 \rightarrow 0$ and $r_1 \rightarrow \infty$. The boundary $\partial M$ of $M$ is  ${\pmb R}  \times (C_0 \cup  C_1)$.

The metric $\sigma$ induced by (\ref{metric})  on a two-sphere of constant  $r$  is $d\sigma^2 = r^2 d\theta^2 + r^2 \sin^2\theta d \phi^2$.  The three-dimensional vector $q$ normal to the sphere is $q = \sqrt{1-2m(r)/r} \frac{\partial}{\partial r}$.

The  action (\ref{acsg2}) for these solutions is a sum of four terms
\begin{eqnarray}
I = I_1 +I_2 +I_3 +I_4,
\end{eqnarray}
where
\begin{eqnarray}
I_1 &=& - \left(\int dt\right) \frac{1}{8 \pi}\int_\Sigma d^3x \sqrt{h}   \nabla_i\nabla^iL \\
I_2 &=&  - \left(\int dt\right) T_{\infty} S_{mat}
\\
I_3 &=&  \frac{1}{8 \pi}\left(\int dt\right)  \int_{C_1} d^2s  L \sqrt{\sigma} (K - K_1) \\
I_4 &=&  - \frac{1}{8 \pi} \left(\int dt\right)  \int_{ C_0} d^2s L\sqrt{\sigma} (K - K_0),
\end{eqnarray}
where $s$ denotes  coordinates on the two-sphere,
 \begin{eqnarray}
 K = \sqrt{1 - \frac{2m}{r}} \left( - \frac{1}{L}\frac{dL}{dr} + \frac{2}{r}\right), \label{2r}
 \end{eqnarray}
$K_0$ and $K_1$ are the values of $K_{ref}$ near the surfaces ${\pmb R} \times C_0$ and ${\pmb R} \times C_1$, respectively.  We also used the fact that $\sqrt{\gamma} = L \sqrt{\sigma}$.

By Gauss' law the integral in $I_1$ equals $ \int_{C_1} d^2 s q^i \nabla_iL -  \int_{C_0} d^2 x q^i \nabla_iL  $. Using the asymptotic expressions  $L =\eta/\sqrt{r}$ for $r \rightarrow 0$ and $L = \sqrt{1-2M/r}$ for $r \rightarrow \infty$, we obtain
\begin{eqnarray}
I_1 =  - \left(\int dt\right) \left(\frac{1}{2} M + \frac{1}{4} \sqrt{2M_0}\eta\right)
\end{eqnarray}
The standard choice for  $K_1$ in $I_3$ is $K_1 = 2L/r$ \cite{GiHa77}, so that $K - K_1 = - dL/dr $ as $r \rightarrow \infty$.
 Then,
\begin{eqnarray}
I_3 = -\frac{1}{2}  \left(\int dt\right) M.
\end{eqnarray}

The part of $I_4$ that involves the integral of $K$ is finite and equal to $- \frac{5}{4} (\int dt) \sqrt{2M_0}\eta$. The problem is that there is no natural choice for $K_0$ in $I_4$. The usual prescription  is that $K_0$ is defined by the embedding of the bounding surface into flat three space. But for $r = 0$, $L$ diverges. Hence, flat space near $r = 0$ corresponds to a spacetime metric with a curvature singularity, which a highly unnatural choice for a reference configuration.

In principle,
  $K_0$ should be derived from a fundamental theory for the renormalization of the gravitational path-integral. Nonetheless, its general form near $r = 0$ can be determined by a simple consistency argument.

First,    $K_0$ contributes finitely to $I_4$ only if it  diverges like $r^{-3/2}$ near the singularity. Furthermore,  $K_0$ must vanish for regular solutions, so it should be proportional to some power of $M_0$.  Since $K_0$ is interpreted as  extrinsic curvature,  it has dimensions of inverse length. $K_0$ is compatible with the above conditions, only if
\begin{eqnarray}
K_0 = c \frac{\sqrt{2M_0}}{r^{3/2}},
\end{eqnarray}
for some dimensionless constant $c$ that remains undetermined.

It follows that
 \begin{eqnarray}
 I_4 = - \frac{5-2c }{4} \left(\int dt \right) \sqrt{2M_0} \eta.
 \end{eqnarray}
Consequently,
\begin{eqnarray}
I = - \left(\int dt\right) \left( M +T_{\infty} S_{mat} + \lambda  \sqrt{2M_0} \eta \right). \label{Icl}
\end{eqnarray}
where we set $\lambda = \frac{1}{2}(3-c)$.

Euclidean Quantum Gravity suggests a fundamental relation between the Helmholtz free energy associated to a static solution of Einstein's equations and the value of the Euclidean action for this solution. Given a static solution $(g_0, \zeta_0)$ to Einstein's equations, we define the associated Euclidean metric $g_0^E$  by substituting the Killing time $t$ with $-i \tau$, where $\tau$ is a periodic coordinate with period $\beta$.
 Then, the Helmholtz free energy $F$ associated to the solution $(g_0, \zeta_0) $ is
\begin{eqnarray}
\beta F = I_E(g^E_0, \zeta_0) , \label{euclidf}
\end{eqnarray}
where $I_E = -i I$ is the Euclidean action.

Using Eq.(\ref{Icl}) to Eq.  (\ref{euclidf}), we find
\begin{eqnarray}
F = M +T_{\infty} S_{mat} + \lambda \sqrt{2M_0} \eta. \label{helmh}
 \end{eqnarray}

\section{Singularity stars for other EoS }
In this section, we reproduce the results of the main text for several different EoS.
\subsection{Interacting fermion gas}

First, we express Eqs. (\ref{nff}---\ref{rff}) for the number density $n$, the pressure $P$ and the energy density $\rho$ as $n = D\mu^{-1} \nu_0(Y), P = D \upsilon_0(Y)$ and $\rho = D u_0(Y)$. Then, we observe that any  EoS of the form
\begin{eqnarray}
n &=& D \mu^{-1} \nu_0(Y) \label{nint}\\
P &=& D[\upsilon_0(Y) - z(Y)]\\
\rho &=& D[u_0(Y) + z(Y)], \label{roint}
\end{eqnarray}
for some function $z(Y)$,  leads  (i)  to $b$-independent structure equations  and (ii) to Eq. (\ref{Omega2}) for the free energy  $\Omega$.

 The mean field theory for a fermion field interacting with a single scalar field through a Yukawa coupling leads to such an equation of state \cite{WC}, with
\begin{eqnarray}
z(Y) = \alpha \left[  Y\sqrt{Y^2-1} - \sinh^{-1}\sqrt{Y^2-1}\right]^2, \label{chi}
\end{eqnarray}
where $\alpha$ is a dimensionless constant proportional to the square of the Yukawa coupling constant. This EoS is derived from a  Lorentz invariant action, through an approximation that preserves Lorentz covariance.

Eq. (\ref{chi}) leads to a well defined EoS for $\alpha <  \alpha_{max} = \frac{2}{3}$. For $\alpha >\alpha_{max}$,  the pressure becomes negative. In Fig. 6, we plot the spiral curve of regular solutions for $\alpha =0$ and for $\alpha = 0.65$ (a value very close to $\alpha_{max}$). The solutions almost coincide on the segment of the curve between the origin and $w_{max}$, but they differ significantly at the  metastable configurations.

Next, we   apply the MEP to solutions of the TOV equation  for $\alpha = 0.65$. The corresponding Oppenheimer-Volkoff limit is $w_{max} \simeq 0.155$. Again, we find that for $w < w_{max}$, the MEP identifies regular solutions as equilibrium states and that for $w > w_{max}$, equilibrium states are singular---see, Fig. 7. In Fig. 8, we plot  the equilibrium entropy as a function of $w_B$, and we find that the ratio $w_-/w_{max}$ is about 2.7.

\begin{figure}[]
\includegraphics[height=5cm]{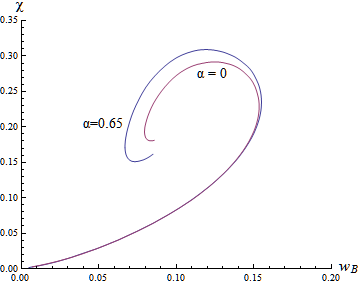} \caption{ \small  The compactness parameter $\chi = 2w_B/x_B$ as a function of the dimensionless mass $w_B$ for regular solutions characterized by the EoS (\ref{nint}---\ref{chi}).  The two curves shown corresponds to $\alpha = 0$ and to $\alpha = 0.65$. }
\end{figure}

  \begin{figure}[]
\includegraphics[height=7.4cm]{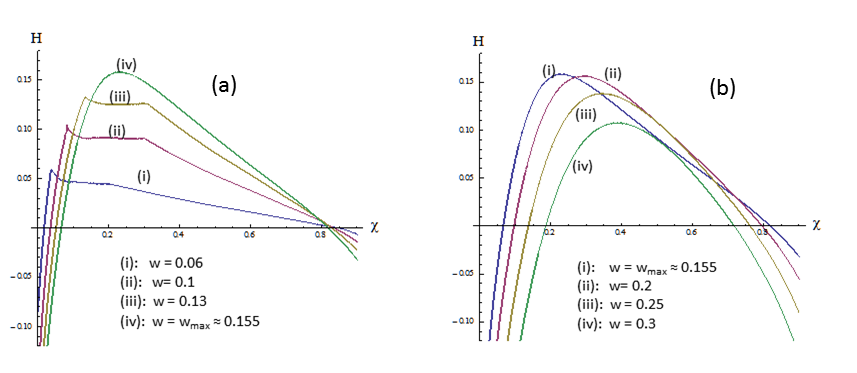} \caption{ \small  The free  entropy $H(w_b, x_B)$  of Eq. (\ref{Hf}) as a function of $\chi = 2w_B/x_B$ for constant $w_B$, for the EoS  (\ref{nint}---\ref{chi}).  In Fig. 7.a,  where $w_B \leq w_{max}$,  global maxima correspond to regular solutions.   In Fig. 7.b, $w_B \geq w_{max}$, and the global maxima correspond to singular solutions.
}
\end{figure}
\begin{figure}[]
\includegraphics[height=4cm]{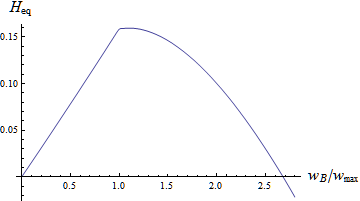} \caption{ \small  The free entropy $H_{eq}$ of the equilibrium solutions  as a function of the ratio $w_B/w_{max}$ for the EoS  (\ref{nint}---\ref{chi}).
}
\end{figure}

\subsection{Polytropic EoS}
Next, we consider matter described by Gratton's EoS \cite{Gra64}
\begin{eqnarray}
\rho = A P^{\gamma} + 3 P \label{gratton},
\end{eqnarray}
where $A$ is a positive constant and $0 < \gamma < 1$. The EoS (\ref{gratton}) is polytropic at low densities; $\gamma = \frac{n}{n+1}$ where $n$ is the standard  polytropic index. For large $\rho$, the EoS describes
ultra-relativistic particles. Note that, in general, $A$ may be a function of $b$.

We define $\rho = \rho_0 u$ and $P = \rho_0 \upsilon$ with  $\rho_0 = A^{\frac{1}{1-\gamma}}$. Then, the EoS (\ref{gratton}) becomes
\begin{eqnarray}
u = \upsilon^{\gamma} + 3 \upsilon.
\end{eqnarray}

By Eq. (\ref{dpt}), $\upsilon = \left[ \frac{(T/T_0)^{4(1-\gamma)}-1}{4}\right]^{\frac{1}{1-\gamma}}$ for some constant $T_0$. Pressure is well-defined only for $T> T_0$. The EoS  is thermodynamically consistent only if  $\upsilon$ and $u$ vanish for $T  < T_0$. Thus, we express the EoS (\ref{gratton}) as
\begin{eqnarray}
t&>& 1:  u =  \frac{1}{4}(3t^{4(1-\gamma)} +1) \left[ \frac{t^{4(1-\gamma)}-1}{4}\right]^{\frac{\gamma}{1-\gamma}} , \upsilon =  \left[ \frac{t^{4(1-\gamma)}-1}{4}\right]^{\frac{1}{1-\gamma}}\nonumber\\
t&<& 1: u = \upsilon = 0, \label{polyeos}
\end{eqnarray}
where $t  = T/T_0$. (Note that the condition $t < 1$ for the vanishing of the pressure is equivalent to the condition $Y< 1$ of Sec. 3.1, if we identify $T_0 = \mu/b$, where $\mu$ is the neutron mass. Then $t = b T/\mu$.)

We define the dimensionless variables $x = \sqrt{4 \pi \rho_0} r$ and $w = \sqrt{4 \pi \rho_0} m$. Then, Eqs. (\ref{TOV}---\ref{dN}) become
\begin{eqnarray}
\frac{dw}{dx} &=& x^2u(t) \label{wx3}\\
\frac{dt}{dx} &=& - \frac{t (w+x^3 \upsilon(t))}{x^2(1-\frac{2w}{x})} \label{tx3}
\end{eqnarray}

We integrate the equations from the boundary $x = x_B$ inwards with boundary conditions $w(x_B) = w_B$ and $t(x_B) = 1$.
For singular solutions, $t$ becomes unity at a finite distance $x = x_0$ from the singularity, so that for $x < x_0$, $t =  \sqrt{\frac{1+2w_0/x_0}{1+2w_0/x}}$. Then, we evaluate the singularity entropy
\begin{eqnarray}
S_{sing} = -  \frac{4}{T_0 \sqrt{4\pi \rho_0}} \frac{w_0}{\sqrt{1+\frac{2w_0}{x_0}}}.
\end{eqnarray}

The free entropy (\ref{ometot}) becomes $\Omega = \frac{1}{T_0 \sqrt{4\pi \rho_0}} H$, where
\begin{eqnarray}
H =  \left[  \int_{x_0}^{x_B}  \frac{ t^{3-4\gamma}  x^2dx}{ \sqrt{1 - \frac{2w}{x}}}\left[ \frac{t^{4(1-\gamma)}-1}{4}\right]^{\frac{\gamma}{1-\gamma}}  - 4  \frac{w_0}{\sqrt{1+\frac{2w_0}{x_0}}}\right]. \label{Omega3}
\end{eqnarray}

For $\gamma = \frac{3}{5}$, the EoS (\ref{polyeos}) at small $\rho$ describes  non-relativistic free fermions. There is little qualitative difference from the EoS of free relativistic particles. We focus on smaller values of $\gamma$ that lead to to regular solutions with higher compactness parameters. In Fig. 9, we plot the spirals of the regular solutions for $\gamma = \frac{1}{2}$ and $\gamma = \frac{1}{3}$. The maximum masses are $0.435$ and $0.265$, respectively. The application of the MEP leads to the same results:   equilibrium states are regular solutions for $w \leq w_{max}$ and singular solutions for    $w > w_{max}$---see, Fig. 10. As shown in Fig. 11, the ratio $w_-/w_{max}$ is about $2.5$ for both values of $\gamma$.

\begin{figure}[]\label{polyreg}
\includegraphics[height=6cm]{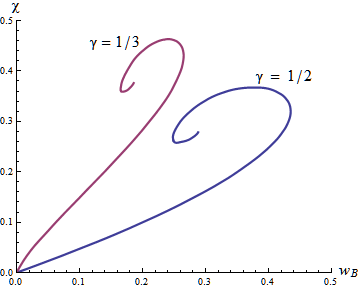} \caption{ \small  The compactness parameter $\chi = 2w_B/x_B$ as a function of the dimensionless mass $w_B$ for regular solutions with the EoS (\ref{polyeos}).  }
\end{figure}

 \begin{figure}[]
\includegraphics[height=13cm]{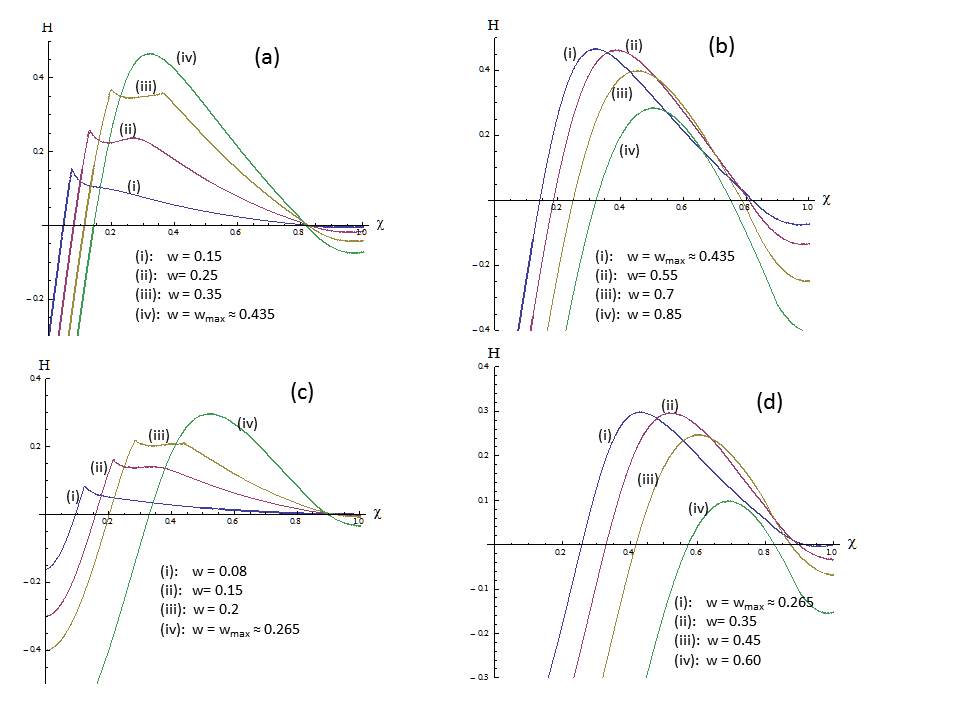} \caption{ \small  The free entropy $H(w_B, x_B)$ for the EoS (\ref{polyeos}) as a function of $\chi = 2w_B/x_B$. In Figs. 10.a and 10.b, $\gamma = \frac{1}{2}$; In Figs. 10.c and 10.d, $\gamma = \frac{1}{2}$.
In both cases, global maxima correspond to regular solutions if
    $w_B \leq w_{max}$ and to singular solutions if  $w_B > w_{max}$.
}
\end{figure}

\begin{figure}[]
\includegraphics[height=4cm]{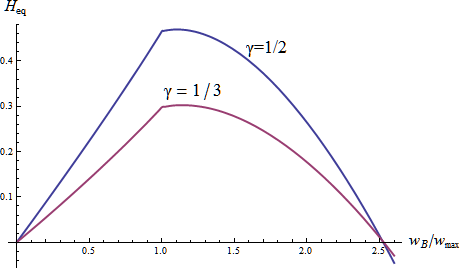} \caption{ \small  The free entropy $H_{eq}$ of the equilibrium solutions for the EoS (\ref{polyeos}) with different values of $\gamma$ as a function of the ratio $w_B/w_{max}$.}
\end{figure}

 \subsection{MIT bag EoS}
Next, we consider a toy EoS for quark matter that follows from  an elementary version of the MIT bag model \cite{MIT}. The model assumes massless and non-interacting quarks confined in  a `bag' that originates from the QCD vacuum. The confinement is described by the `bag constant' ${\cal B}$ that determines the energy difference between the standard and the QCD vacua.

The EoS is
\begin{eqnarray}
n &=& C_2(b) T^3 \label{nmit}\\
\rho &=& C_3(b) T^4 +{\cal B}\\
P &=& \frac{1}{3}   C_3(b) T^4  - {\cal B},\label{pmit}
\end{eqnarray}
 where
 \begin{eqnarray}
 C_n(b) = \frac{g}{\pi^2} \int_0^{\infty} \frac{x^n}{e^{x-b} + 1},
 \end{eqnarray}
and  $g$ is the number of independent quark species. Eqs. (\ref{nmit}---\ref{pmit}) imply that $P = \frac{1}{3}(\rho - \rho_0)$, where $\rho_0 = \frac{4}{3} {\cal B}$. Positivity of pressure implies that $n$ vanishes for $T < T_0 := 3{\cal B}/C_3(b)$.

We write $\rho = \frac{\rho_0}{4} u(t)$ and $P = \frac{\rho_0}{4}\upsilon(t)$ in terms of
 the dimensionless temperature $t = T/T_0$. Then, the EoS reads
\begin{eqnarray}
t&>& 1:  u =  3 t^4 + 1,
\upsilon = t^4 - 1\nonumber\\
t&<& 1: u = \upsilon = 0. \label{miteos}
\end{eqnarray}

The boundary of the star is defined by $\rho(r_B) =  \rho_0$. This implies that $T(r_B) = T_0$ and the temperature at infinity is $T_{\infty} = T_0\sqrt{1 -\frac{2M}{r_B}}$.

We define the dimensionless variables $x = \sqrt{ \pi \rho_0} r$ and $w = \sqrt{ \pi \rho_0} m$. Then, Eqs. (\ref{TOV}---\ref{dN}) become identical to Eqs. (\ref{wx3}---\ref{tx3}).

We integrate the equations from the boundary $x = x_B$ inwards with boundary conditions $w(x_B) = w_B$ and $t(x_B) = 1$.
For singular solutions, $t$ becomes unity at a finite distance $x = x_0$ from the singularity, so that for $x < x_0$, $t =  \sqrt{\frac{1+2w_0/x_0}{1+2w_0/x}}$.
The free entropy (\ref{ometot}) becomes $\Omega = \frac{1}{T_0 \sqrt{\pi \rho_0}} H$, where
\begin{eqnarray}
H =  \left[ 4 \int_{x_0}^{x_B}  \frac{ t^3 x^2dx}{ \sqrt{1 - \frac{2w(x)}{x}}} - 4  \frac{w_0}{\sqrt{1+\frac{2w_0}{x_0}}}\right]. \label{Omega3s}
\end{eqnarray}

 In Fig. 12, we plot the spiral curve of regular solutions. The maximum mass is found at $w_{max} \simeq 0.092$. Again, we find that for $w < w_{max}$, the MEP identifies regular solutions as equilibrium states and that for $w > w_{max}$, equilibrium states are singular---see Fig. 13. In Fig. 14, we plot the free  entropy of the equilibrium solutions as a function of $w_B/w_{max}$ and we find    $w_-/w_{max} \simeq 2.7$.

\begin{figure}[]
\includegraphics[height=6cm]{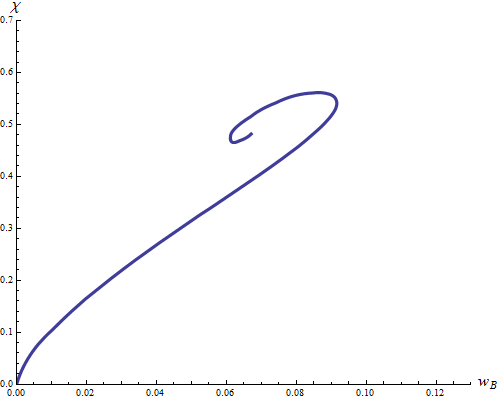} \caption{ \small  The compactness parameter $\chi = 2w_B/x_B$ as a function of the dimensionless mass $w_B$ for regular solutions with the EoS (\ref{miteos}). }
\end{figure}

  \begin{figure}[]
\includegraphics[height=6.3cm]{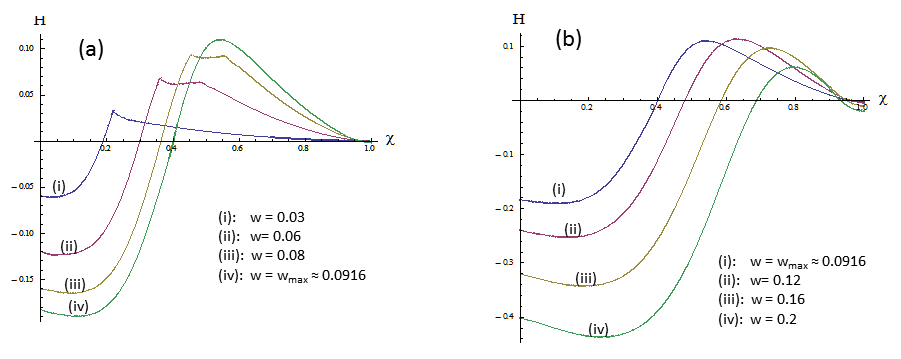} \caption{ \small  The free entropy $H(w_B, x_B)$  for the EoS (\ref{miteos}) as a function of $\chi = 2w_B/x_B$ for constant $w_B$.  In Fig. 13.a,    $w_B \leq w_{max}$,  global maxima correspond to regular solutions.  In Fig. 13.b, $w_B \geq w_{max}$, all  global maxima correspond to singular solutions.
}
\end{figure}

\begin{figure}[]
\includegraphics[height=4cm]{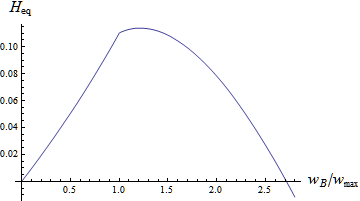} \caption{ \small  The free entropy $H_{eq}$ of the equilibrium solutions  for the EoS (\ref{miteos}) as a function of the ratio $w_B/w_{max}$.}
\end{figure}

\subsection{The asymptotic behavior of the EoS}
Finding  the conditions that an EoS must satisfy in order to be compatible with relativity is an old problem that has not yet been resolved.  The causality condition $|\partial P/\partial \rho| \leq 1$ is well accepted, since it  guarantees that the speed of sound on the material never exceeds the speed of light.  Other conditions have been suggested but they are not universally accepted. For example, Landau and Lifschitz proposed \cite{LL} that
\begin{eqnarray}
P \leq \frac{1}{3} \rho. \label{LL}
\end{eqnarray}
This follows from the assumption that for sufficiently high energy all particles behave like massless particles even in presence of interactions.
 Counterexamples exist \cite{Zeld, GH}; however,   theories characterized by asymptotic freedom are expected to satisfy (\ref{LL}) for $\rho \rightarrow \infty$.

Relativistic constraints to the EoS, other than causality, do not usually enter the compact-star models. Potential violations would typically appear at very high densities, well beyond the regime of physical interest. However, here we consider {\em virtual} solutions to the Einstein equations at extreme densities and far from thermodynamic equilibrium. In this regime,  the asymptotic behavior of the EoS as $\rho \rightarrow \infty$  is relevant.

In all examples  that we have considered so far, the EoS satisfies Eq. (\ref{LL}).  The free entropy $\Omega$ remains finite  as $\chi \rightarrow 1$ for fixed $M$, the values of $\Omega$ being much smaller than the global maximum. However,  EoS with different asymptotic behavior leads  to
 $\Omega$ blowing up as $\chi \rightarrow 1$. The equilibrium state corresponds to  $\chi = 1$. Then, regular solutions are metastable states rather than equilibrium states. This is the case, for example,  for the EoS of the non-relativistic free fermion gas with a causality-violating  asymptotic behavior, $P \rightarrow \rho^{5/3}$.

The EoS $P = \frac{1}{a}(\rho - \rho_0)$ for  constants $\rho_0$  and   $a \geq 1$ does not violate causality. For $a = 3$, it  describes the MIT bag model; for $a = 1$, it coincides with the EoS for the `most compact' star considered in Ref. \cite{KSF}.

 As in the model of Sec. E.3, a transition temperature $T_0$ appears as an integration constant. Writing $t = t/T$,
 $\rho =  \frac{\rho_0}{1+a} u(t)$,  and  $P = \frac{\rho_0}{1+a} \upsilon(t)$, the EoS becomes

\begin{eqnarray}
t &>& 1:  u = a t^{1+a} + 1,
\upsilon = t^{1+a} - 1 \\
t&<& 1 : u = \upsilon = 0.
\end{eqnarray}

We define the dimensionless variables $x = \sqrt{ 4\pi \rho_0/(a+1)} r$,  and $w = \sqrt{ \pi \rho_0/(a+1)} m$ Then, Eqs. (\ref{TOV}---\ref{dN}) coincide with Eqs. (\ref{wx3}, \ref{tx3}), and the free entropy becomes
$\Omega = \frac{1}{T_0 \sqrt{4\pi \rho_0/(a+1)}} H$, where
\begin{eqnarray}
H =  \left[ (a+1) \int_{x_0}^{x_B}  \frac{ t^a x^2dx}{ \sqrt{1 - \frac{2w}{x}}} - 4  \frac{w_0}{\sqrt{1+\frac{2w_0}{x_0}}}\right]. \label{Omega4}
\end{eqnarray}

The implementation of the MEP for $a = 1$ is shown in Fig. 15. Regular solutions still define  local maxima of the free entropy, but these maxima are no longer global, because the free entropy diverges as $\chi \rightarrow 1$. Matter with this EoS cannot form stable compact stars, because they have always less entropy from the black-hole-like configurations at $\chi = 1$. We find that this behavior persists for  values of $a < 3$. For $a \geq 3$, we recover the results presented in the main text and in the previous sections.

  \begin{figure}[]
\includegraphics[height=8cm]{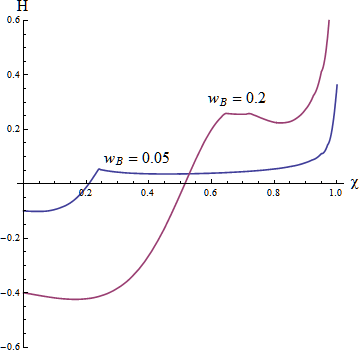} \caption{ \small  The free entropy $H(w_B, x_B)$  of Eq. (\ref{Omega4}) for $a = 1$ as a function of $\chi = 2w_B/x_B$ for constant $w_B$. Regular solutions correspond to local maxima of the free entropy; however, they are not global maxima, because the free entropy diverges as $\chi \rightarrow 1$.
}
\end{figure}

There are two possible interpretations of this result. Either relativistic consistency implies Eq. (\ref{LL})   as $\rho \rightarrow \infty$ (a result that remains to be proven from first principles), or matter that violates (\ref{LL}) exists but it cannot form stable compact stars.

Finally, we note that any EoS that is employed for neutron stars  can also be employed for the study of singularity stars, as long as a high density cut-off $\rho_c$ is included so that
  Eq. (\ref{LL}) is enforced  for all $\rho > \rho_c$.

\end{document}